\documentclass[a4paper,11pt]{article}

\pdfoutput=1
\usepackage[totalwidth=480pt,totalheight=680pt]{geometry}
\usepackage{amsmath,amssymb,amsfonts,amsthm,mathrsfs,bbm,time}
\usepackage{graphicx}
\usepackage{color}

\usepackage{natbib}
\bibpunct{(}{)}{,}{a}{,}{,}
\usepackage{ifthen}
\newboolean{preview}
\setboolean{preview}{true}  
\ifthenelse{\not\boolean{preview}}{%
\usepackage[notref,notcite]{showkeys}
\usepackage{fancyheadings}
\pagestyle{fancyplain}

\lhead{\leftmark}
\rhead{\thepage}
\cfoot{{\tt \RCSId}}
}{}

\newcommand{\E}{\mathbb{E}}

\newcommand{\1}{{\mathbbm{1}}}

\begin{document}
\parskip
1ex 


\title{On variance stabilisation by double Rao-Blackwellisation\footnote{This 
    work has been supported by the Agence Nationale de la Recherche (ANR, 212, rue de Bercy 75012 Paris)
    through the 2006-2008 project {\sf Adap'MC}. The third author is particularly grateful to 
    Andrew Barron for his suggestion to look at biased datasets in the context of mixtures, 
    during the Bayesian Nonparametric Workshop in Roma, June 2004.
    }}

\author{{\sc Alessandra Iacobucci,}\\
  {\em CEREMADE, Universit\'e Paris Dauphine \&\ CNRS, Paris}\\
  {\sc Jean-Michel Marin,}\\
  {\em INRIA \textsc{Saclay}, Project \textsc{select}, Universit\'e Paris-Sud, \&\ CREST, INSEE}\\
  \&\ {\sc Christian P. Robert}\\
  {\em CEREMADE, Universit\'e Paris Dauphine \&\ CREST, INSEE}}

\maketitle

\begin{abstract}
  Population Monte Carlo has been introduced as a sequential importance sampling technique
  to overcome poor fit of the importance function. In this paper, we compare the performances of
  the original Population Monte Carlo algorithm with a modified version that eliminates the influence
  of the transition particle via a double Rao-Blackwellisation.
  This modification is shown to improve the exploration of the modes through an large
  simulation experiment on posterior distributions of mean mixtures of distributions.

  \noindent{\bf Keywords:} Importance Sampling mixture, adaptive Monte Carlo, Population Monte
  Carlo, multimodality, mixture of distributions, random walk.
\end{abstract}

\section{Introduction}
When \cite{cappe:guillin:marin:robert:2003} introduced Population Monte Carlo (or PMC) as
a repeated Sampling Importance Resampling procedure, the purpose was to overcome the
shortcomings of Importance Sampling (abbreviated to IS in the following) procedures, 
while preserving the advantages of IS compared to
alternatives such as Markov Chain Monte Carlo methods,
namely the possibility of developing parallel implementations, 
which becomes more and more important with the
generalisation of multiple core machines and computer clusters,
and of allowing for easy assessment of the Monte Carlo error and,
correlatively, for the development of on-line calibration mechanisms.

Indeed, adaptive Monte Carlo naturally answers the difficulties of finding a 
``good" importance function in IS by gradually improving this importance function
against a given target density, based on some form of Monte Carlo approximation.
Previous simulations are used \citep[see, e.g.,][]{douc:guillin:marin:robert:2005,
douc:guillin:marin:robert:2007b} to modify proposal distributions represented as
mixtures of {\em fixed} proposals in order to increase the weights of the most 
appropriate components. When the proposal is inspired from random walk structures 
as in Metropolis--Hastings algorithms and when the update of those weights is too crude, 
as in \cite{cappe:guillin:marin:robert:2003}, the improvement is so short-sighted that
multiple iterations do not increase the efficiency of the method, unless a Rao-Blackwell
step is added to replace the actual proposal
with the mixture proposal in the importance àweight \citep[see][]{douc:guillin:marin:robert:2005}.
More recently, \cite{cappe:douc:guillin:marin:robert:2007} have thus replaced random walk
proposals based on earlier samples with a new PMC scheme with parameterised proposals whose
parameters are estimated from earlier samples, leading to a monotonic decrease in the entropic
distance to the target distribution. There is indeed a myopic feature in random walk proposals,
namely that, once a (first) sample has exhibited some high density regions for the target
distribution, the algorithm is reluctant to allow for a wider exploration of the
support of the target and it may thus miss important density regions, missing the energy
to reach forward to these other regions of  interest.

In this paper, we nonetheless focus on random walk proposals because those kernels are more open to complex
settings than on independent proposals---the later indeed require some preliminary knowledge about 
the target or else an major upgrade in computing power to face a much larger number of components 
in the mixture. More precisely, we present an experimental assessment of the use of a 
so-called {\em double Rao-Blackwellisation scheme} towards a better exploration of the modes of the target distribution.
The {\em second} Rao-Blackwellisation step used in this {\em double} Rao-Blackwellisation scheme is 
essentially an integration over the particles of the previous sample used 
in the random walk proposal. While this leads to an additional computing cost in the derivation 
of the importance weights, double Rao-Blackwellisation undoubtedly brings a significant improvement 
in the stability of those importance weights and therefore justifies (in principle) the use of this 
additional step. Nevertheless, since an analytic proof of this improvement brought by double Rao-Blackwellisation
is too delicate to contemplate, we use an intensive Monte Carlo experiment to establish the clear
improvement in the case of a posterior distribution associated with a Gaussian location mixture 
and a sample with outliers---in order to increase the number of modes.

The paper is organised as follows: In Section \ref{sec:pmc},
we recall the basics of our population Monte Carlo algorithm and we introduce the
double Rao-Blackwellisation modification.
In Section \ref{sec:mmix}, we motivate the choice of Gaussian mean mixtures 
as a worthy Monte Carlo experiment to test the mode finding abilities of both
algorithm. Section \ref{sec:Hcore} describes the implementation of the Monte
Carlo experiment and describes the results. Section \ref{sec:concq} contains
a short discussion.

\section{Population Monte Carlo}\label{sec:pmc}

Given a target distribution with density $\pi$ that is known up to a normalising constant, 
the grand scheme of PMC is the same as with other Monte Carlo---including MCMC---methods, 
namely to produce a sample that is distributed from $\pi$ without resorting to direct
simulation from $\pi$.  Let us recall that, once a sample $(X_1,\dots,X_N)$ is produced
by sampling importance resampling \citep[see, e.g.,][]{robert:casella:2004,marin:robert:2007}, i.e.,
by first simulating $X_i\sim f(x)$, second producing importance weights
        $\omega_i\propto \pi(X_i)/f(X_i)$,
and third resampling the $X_i$'s by multinomial/bootstrap sampling with weights $\omega_i$,
this SIR sample provides an approximation to the target distribution $\pi$ and can be
used as a stepping stone towards a better approximation to $\pi$.

\subsection{PMC basics}\label{sub:pmc.1.0}
In the PMC algorithm of \cite{cappe:guillin:marin:robert:2003}, if
$$
(X_1,\dots,X_N)
$$
is a sample approximately distributed from $\pi$, it is modified stochastically using
an arbitrary Markov transition kernel $q(x,x')$ so as to produce a new sample
$$
(X_1',\dots,X'_N)
$$
as $X_i'\sim q(X_i,x)$. Conducting a resampling step based on the IS weights
$\omega_i=\pi(X_i')/q(X_i,X_i')$, we then produce a new sample
$$
(\tilde X_1,\dots,\tilde X_N)
$$
that equally constitutes an approximation to the target distribution $\pi$. It is however
necessary to stress that, as established in \cite{douc:guillin:marin:robert:2007b},
repeating this scheme in an iterative manner is only of interest if samples that
have been previously simulated are used to update (or adapt) the kernel $q(x,x')$: 
a kernel $q$ that is fixed over iterations does not modify the statistical properties of the samples.

The choice made in \cite{douc:guillin:marin:robert:2007b} of an adaptive
proposal kernel $q$ represented as a mixture of fixed transition kernels $q_d$,
\begin{equation}
\label{eq:genericPMC}
q_\alpha(x,x') = \sum_{d=1}^D\alpha_d q_d(x,x')\,, \qquad \sum_{d=1}^D\alpha_d = 1\,,
\end{equation}
can improve the efficiency of the kernel in terms of deviance (or relative entropy) 
from the target density. To achieve such an improvement, the weights $\alpha_1,\dots,\alpha_D$ 
must be tuned adaptively at each iteration of the PMC algorithm. 

\subsection{Weight updating}
If $\alpha^{t}=\left(\alpha_1^{t},\ldots,\alpha_D^{t}\right)$ denote the mixture weights
at the $t$-th iteration of the algorithm (where $t=1,\ldots,T$), the update of the $\alpha^t$'s
of \cite{douc:guillin:marin:robert:2007b} takes advantage of the latent variable structure that
underlines the mixture representation, resulting in an integrated EM (Expectation-Maximisation)
scheme. In the current setting, the latent variable $Z$ is the standard component indicator in the mixture
\citep[see, e.g.,][]{marin:mengersen:robert:2004},
with values in $\{1,\ldots,D\}$ such that the joint density $f$ of $x'$ and $z$ given $x$ satisfies
$$
f(z)=\alpha_z \quad \text{and} \quad f(x'|z,x)=q_z(x,x') \, .
$$
The updating mechanism for the $\alpha_d$'s then corresponds to setting the new parameter
$\alpha^{t+1}$ equal to
$$
\arg\max_{\alpha}\,
\E_{\pi\times\pi}^{X,X'}\left[\E_{\alpha^t}^Z\left\{\log(\alpha_Z q_Z(X,X')
    )|X,X'\right\}\right] = \arg\max_{\alpha}\,
\E_{\pi\times\pi}^{X,X'}\left[\E_{\alpha^t}^Z\left\{\log( \alpha_Z )|X,X'\right\}\right]\,,
$$
where the inner expectation is computed under the conditional distribution of
$Z$ for the current value of the parameter, $\alpha^t$, i.e.
$$
f(z|x,x',\alpha^t)=\alpha^t_z q_z(x,x') \bigg/ \sum_{d=1}^D \alpha^t_d q_d(x,x')\,,
$$
with solution 
\begin{equation}
\alpha_d^{t+1}=\E_{\pi\times\pi}^{X,X'} \left[f(d|X,X',\alpha^t) \right]\,.
\label{eq:updalpha}
\end{equation}

\subsection{Single Rao--Blackwellisation updates}
Given that \eqref{eq:updalpha} is not available in closed form,
the adaptivity of the proposed procedure is achieved by approximating this
actualising expression based on the sample that has been previously simulated.
The implementation of the standard PMC algorithm relies on
an arbitrary initial proposal $\mu_0$ that produces a pre-initial sample
$$
(X_{1,-1},\ldots,X_{N,-1})\,,
$$
with importance weights $\pi(X_{i,-1})/\mu_0(X_{i,-1})$. This initial sample allows for the
derivation of a sample
$$
(\tilde X_{1,-1},\ldots,\tilde X_{N,-1})
$$
approximately distributed from $\pi$, using importance sampling based on those weights. The algorithm
then picks arbitrary starting weights $\alpha^0_d$ in the $N$-dimensional 
simplex to produce a (truly) initial sample
$$
(X_{1,0},\ldots,X_{N,0})
$$
from the mixture
$$
X_{i,0}\sim\sum_{d=1}^D \alpha^0_d q_d(\tilde X_{i,-1},x)\,.
$$
In the detail of its implementation, the production of this initial sample is 
naturally associated with latent variables $(Z_{i,0})_{1\le i\le N}$ that indicate from which
component $d$ of the mixture the corresponding $X_{i,0}$ has been
generated $(i=1,\ldots,n)$. From this stage, \cite{douc:guillin:marin:robert:2007b} proceed recursively.
Starting at time $t$ from a sample
$$
(X_{1,t},\ldots,X_{N,t})\,,
$$
associated with $(Z_{i,t})_{1\le i\le N}$
and with the current value of the weights $\alpha^{t,N}$, the normalised importance weights of
the sample point $X_{i,t}$ is provided by
\begin{equation}
  \label{eq:mixtureISweights}
  \bar\omega_{i,t}=\frac{\pi(X_{i,t})}{\sum_{d=1}^D\alpha^{t,N}_dq_d(\tilde X_{i,t-1},X_{i,t})}\bigg/
  \sum_{j=1}^N \frac{\pi(X_{j,t})}{\sum_{d=1}^D\alpha^{t,N}_dq_d(\tilde X_{j,t-1},X_{j,t})}\,.
\end{equation}
To approximate the update step~\eqref{eq:updalpha} in practice, an initial possibility is
$$
\alpha^{t+1,N}_d = \sum_{i=1}^N\bar\omega_{i,t}\1\{Z_{i,t} = d\} \,,
$$
where the sum does not need renormalisation because the $\bar\omega_{i,t}$ sum up to $1$ (over $i$).

This choice is however likely to be highly variable when $N$ is small and/or
the number of components $D$ becomes larger. To robustify this update,
\cite{cappe:douc:guillin:marin:robert:2007} introduce a Rao-Blackwellisation step
\citep[see, e.g.,][Section 4.2]{robert:casella:2004} that consists in
replacing the Bernoulli variable $\1\{Z_{i,t} = d\}$ with its conditional expectation 
given $X_{i,t}$ and $\tilde X_{i,t-1}$, that is,
$$
\alpha^{t+1,N}_d = \sum_{i=1}^N\bar\omega_{i,t}f\left(d\left|
        X_{i,t},\tilde X_{i,t-1},\alpha^{t,N}\right.\right)\,.
$$
This replacement does not involve a significant increase in the computational cost of the algorithm,
while both approximations are converging to the solution of \eqref{eq:updalpha} as
$N$ grows to infinity.

\subsection{Double Rao-Blackwellisation}
While \cite{cappe:douc:guillin:marin:robert:2007} showed through two experimental examples
that the above Rao-Black\-wel\-li\-sa\-tion step does provide a significant improvement in the stability
of the PMC weights (and, thus, {\em in fine}, a reduction in the number of iterations), there
remains an extra-source of variation in the importance weights~\eqref{eq:mixtureISweights}, namely
the dependence of $\bar\omega_{i,t}$ on the previous sample point (or particle) $\tilde X_{i,t-1}$. 
Even though this dependence illustrates the fact that $X_{i,t}$ is indeed generated from the mixture
$$
\sum_{d=1}^D\alpha^{t,N}_dq_d(\tilde X_{i,t-1},x)\,,
$$
and is thus providing a correct IS weight, it also shows a conditioning on the result of a (random) 
multinomial sampling in the previous iteration that led to the selection of $\tilde X_{i,t-1}$ 
with probability $\bar\omega_{i,t-1}$. In other words, by de-conditioning, it is also the case 
that the sample point $X_{i,t}$ has been generated from the (integrated) distribution
\begin{equation}\label{eq:pasodoble}
\sum_{j=1}^N \bar\omega_{j,t-1}\,\sum_{d=1}^D\alpha^{t,N}_dq_d(\tilde X_{j,t-1},x)\,,
\end{equation}
when averaging over all multinomial outputs. This double averaging over both the components of
the mixture and the initial sample points is the reason why we call this representation
{\em double Rao--Blackwellisation}.

The appeal of using \eqref{eq:pasodoble} is that
not only does the averaging over all possible sample points provide a most likely stabilisation
of the weights, but it also eliminates a strange feature of the original approach, namely that
two identical values $x_{i,t}$ and $x_{j,t}=x_{i,t}$ could have different importance weights simply
because their conditioning sample values $\tilde X_{i,t-1}$ and $\tilde X_{j,t-1}$ were different.

Naturally, the replacement of the importance sampling distribution in
\eqref{eq:mixtureISweights} by \eqref{eq:pasodoble} has a cost of $\text{O}(N^3)$
compared with the original $\text{O}(N^2)$, but this is often negligible when
compared with the cost of computing $\pi(X_{i,t})$. (We also stress that some
time-saving steps could be taken in order to avoid computing all the $q_d(\tilde X_{j,t-1},X_{i,t})$
by considering first the distance between $X_{j,t-1}$ and $X_{i,t}$ and keeping only close
neighbours within the sum although the increase in computing time in our case did not justify
the filtering.) In the following experiment, the additional cost resulting from the
double Rao--Blackwellisation does not induce a considerable upsurge in the computing time,
even though it is not negligible. We indeed found an increase in the order of three to five times
the original computing time for the same number $N$ of sampling values. This obviously fails
to account for the faster stabilisation of the IS approximation resulting from using 
double Rao--Blackwellisation. Note also that double Rao--Blackwellisation does not remove the
need to sample the $\tilde X_{j,t-1}$'s: indeed, when simulating each new $X_{i,t}$
from \eqref{eq:pasodoble}, we need to first select which $\tilde X_{j,t-1}$ is going to
be conditioned upon, then second determine which component is to be used.

\section{The Gaussian mean mixture benchmark}\label{sec:mmix}
\begin{figure}
\includegraphics[width=8truecm]{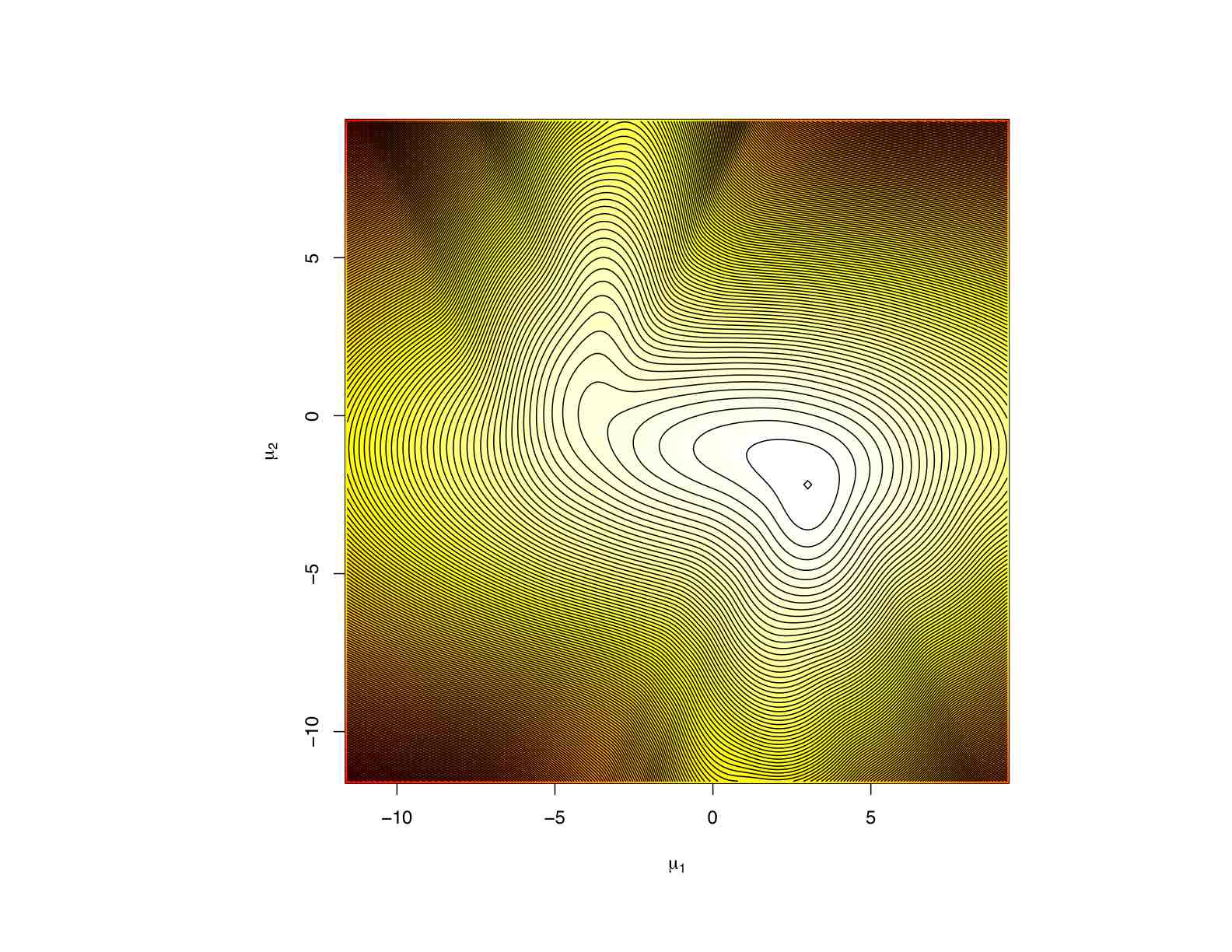} 
\includegraphics[width=8truecm]{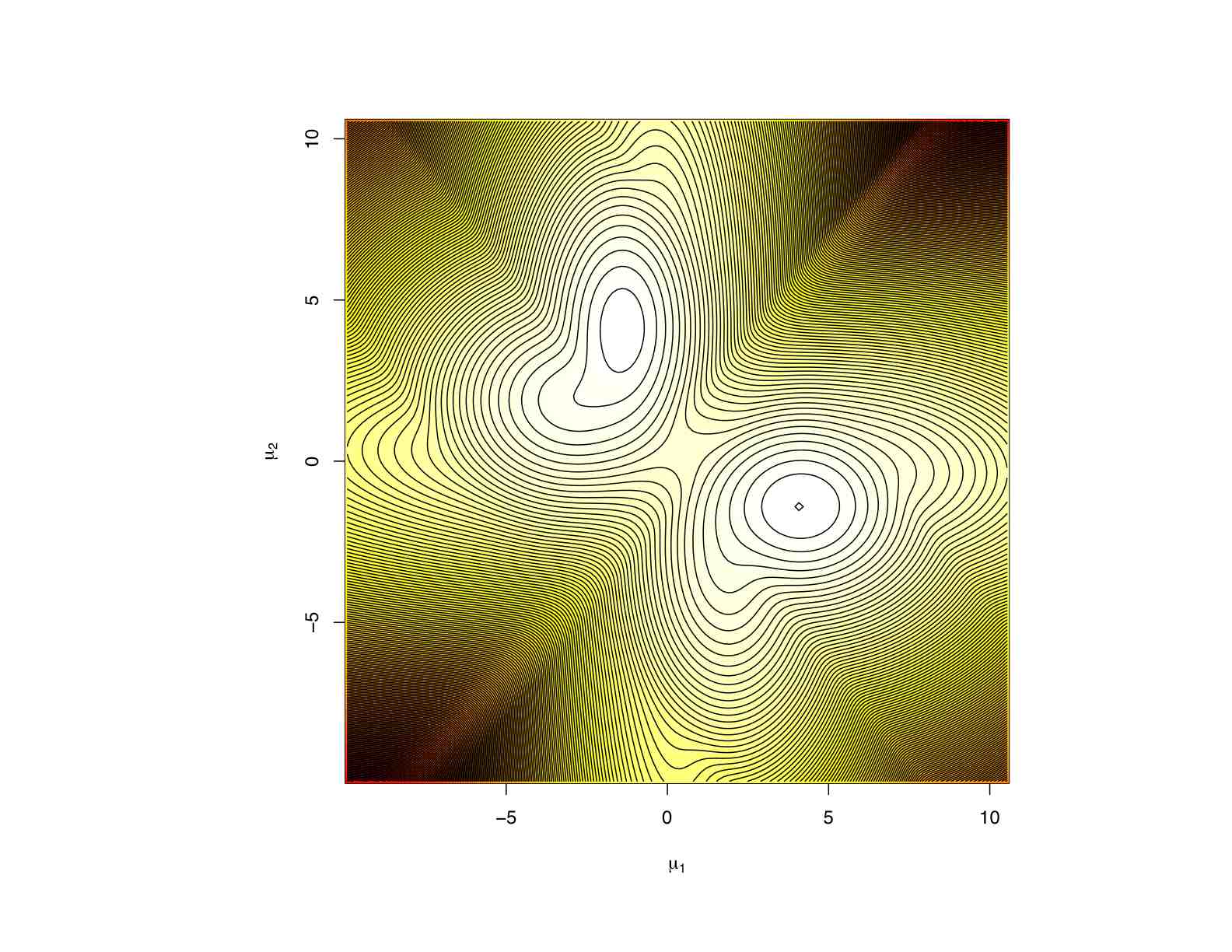} 

\includegraphics[width=8truecm]{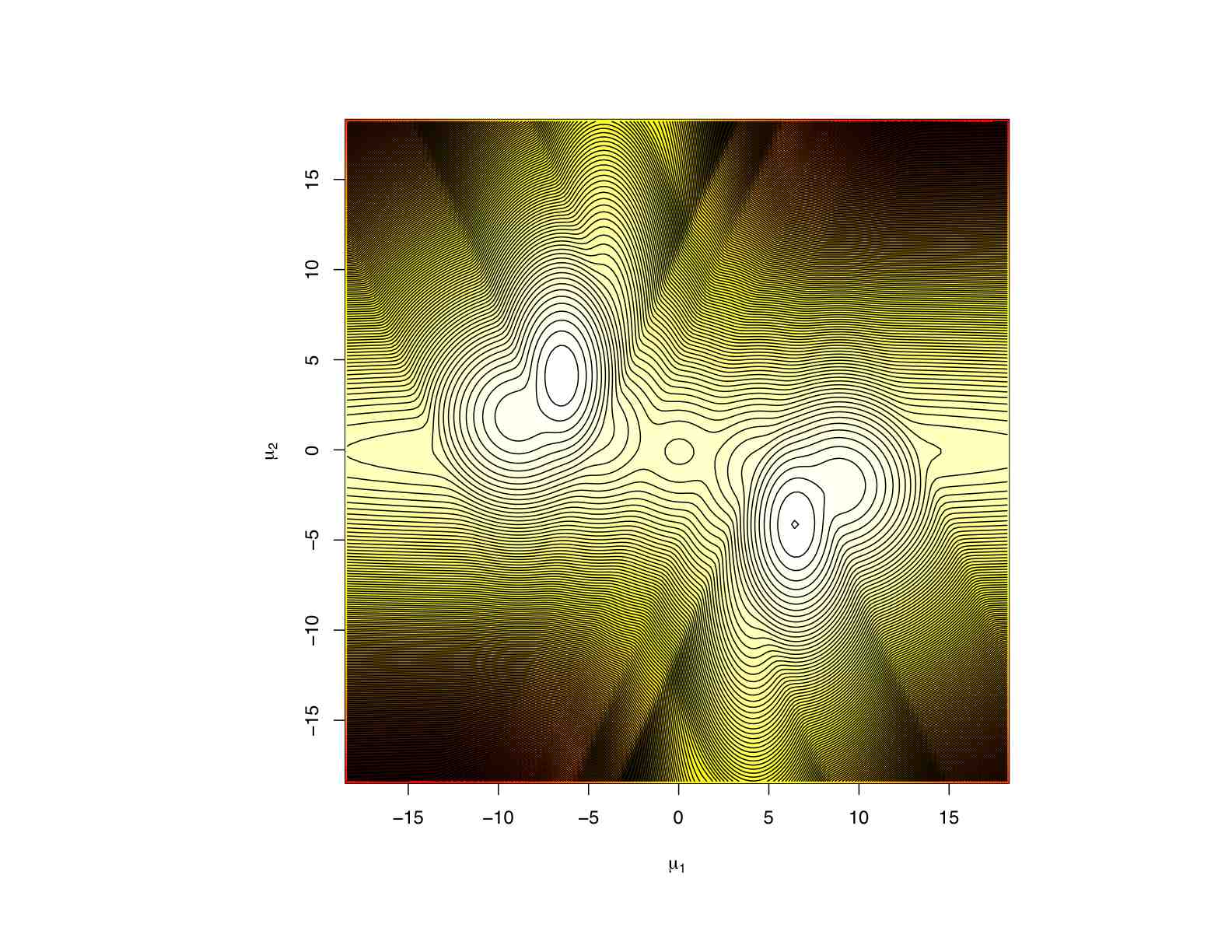} 
\includegraphics[width=8truecm]{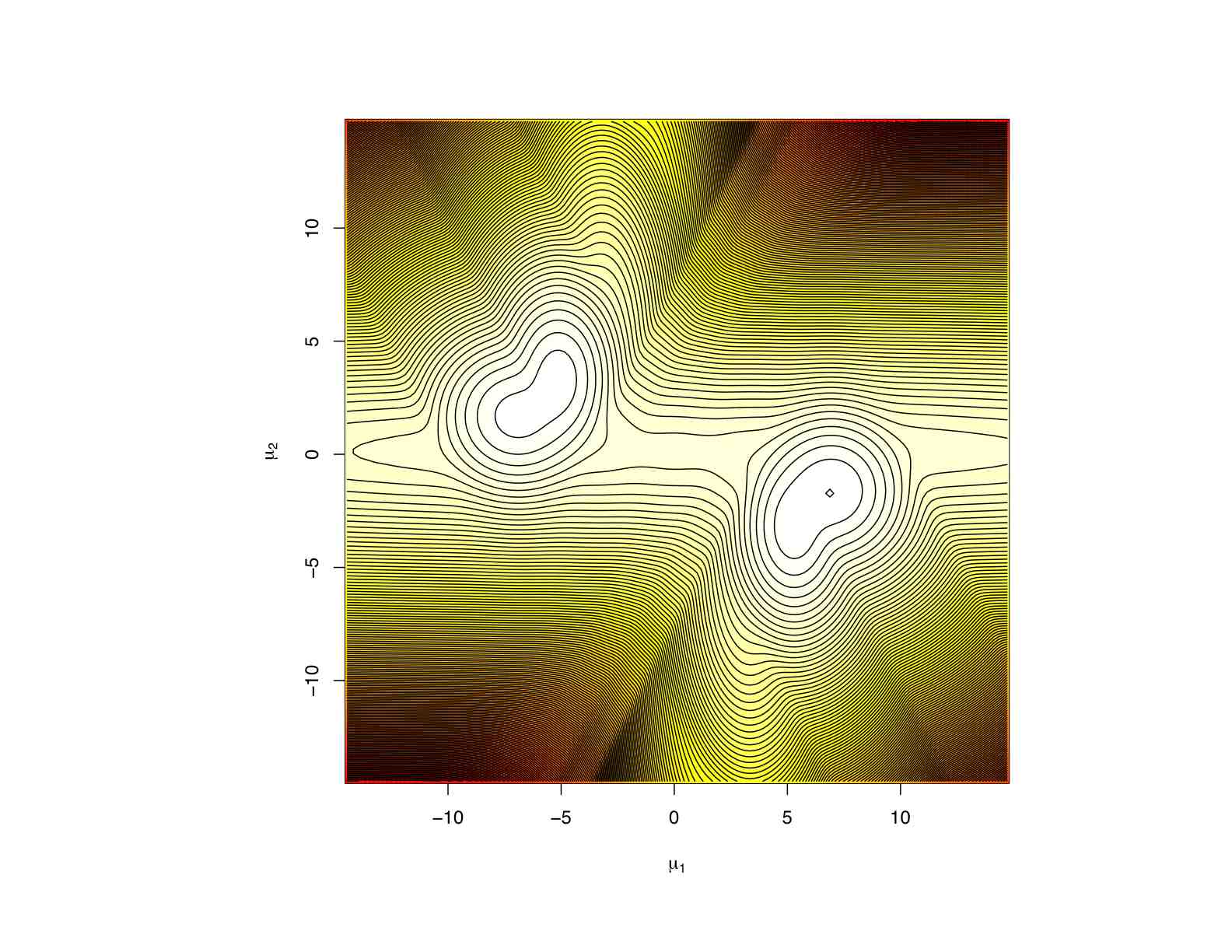} 

\caption{\label{fig:examples}
Series of posterior distributions associated
with \eqref{eq:meanmixture} represented as level sets for various
artificial samples $(x_1,\ldots,x_n)$ and different values of $n$, $\mu_1$, $\mu_2$, $\sigma_1^2$ and $\sigma_2^2$.}
\end{figure}

As benchmark for our Monte Carlo experiment, we consider the case of a one-dimensional Gaussian 
mean mixture distribution, namely
\begin{equation}\label{eq:meanmixture}
x_1,\ldots,x_n \sim p\mathcal{N}(\mu_1,\sigma_1^2)+(1-p)\mathcal{N}(\mu_2,\sigma_1^2)\,,
\end{equation}
with $p$, $\sigma_1^2$ and $\sigma_2^2$ known and $\theta=(\mu_1,\mu_2)$ being the parameter of the model, as in 
\cite{marin:mengersen:robert:2004}. 
Using a flat prior on $\theta$ within a square region, we are thus interested
in simulating from the posterior distribution associated with a given sample
$(x_1,\ldots,x_n)$. The appeal of this example is that it is
sufficiently simple to allow for an explicit characterisation of the attractive
points for the adaptive procedure, being of dimension two, while still illustrating the variety of
situations found in more realistic applications. In particular as already explained in \cite{marin:mengersen:robert:2004},
the model contains 
at least one attractive point that does not correspond to the global minimum of the
entropy criterion as well as some regions of attraction that can eventually
lead to a failure of the algorithm when the data $(x_1,\ldots,x_n)$ is not simulated
from the above mixture but by clumps that induce more modes in the posterior.
Figure \ref{fig:examples} illustrates quite well the diversity of posterior
features when changing the repartition of the sample $(x_1,\ldots,x_n)$. We stress
the fact that those samples, called artificial samples, are not resulting from simulations
from \eqref{eq:meanmixture} but from simulations from a normal mixture with five equal and well-separated components centred in $\mu_1=0$ 
and $\pm \mu_2$, $\pm 2\mu_2$, and with variances $0.1$. This choice was made in order to increase the potential number of modes
on the posterior surface when modelling the data with \eqref{eq:meanmixture}.

\section{Monte Carlo experiment}\label{sec:Hcore}

Given the target distribution defined by \eqref{eq:meanmixture}, we want to study the improvement
brought by the double Rao-Blackwellisation \eqref{eq:pasodoble}
in terms of mode degeneracy, as well as to ascertain the additional
cost of using double Rao-Blackwellisation. We therefore study the performance of
both single and double Rao-Blackwellisation PMC over a whole range of samples, with various
values of $n$ and of the parameters $p$, $\mu_2$ and $\sigma_2$ (since we can always
use $\mu_1=0$ and $\sigma_1=1$ without loss of generality).

\subsection{Automatic mode finding and classification}\label{sub:modfind}
A first difficulty, when building this experiment, is to determine the number of
modes on the posterior surface for the data at hand. We achieve this goal by first
discretising the parameter space in $(\mu_1,\mu_2)$, and then recovering the modal points
by a brute-force search for maxima and saddle-points over the grid and then identifying
their basins of attraction. (The algorithm is based on multidirectional calls to the function 
{\tt turnpoints()} of the {\tt pastecs} package.) 
This is obviously prone to overlook some small local modes but it also allows for a quick 
identification of the modal regions and of their exploration by the PMC algorithm.

Table \ref{tab:modfind} illustrates the results of this procedure for a range of values of
$(n,p,\mu_2,\sigma_2)$. When conditioning upon $(\mu_2,\sigma_2)$ the number of modes is
increasing in $\mu_2$ (since the simulated samples include five clusters that are better
separated) and slightly decreasing in $\sigma_2$ (for apparently the same reason). A similar
table varying upon the pair $(n,p)$ does not show much variation in the number of modes (which
does make sense, since only the relative magnitudes of the modes are changing).

\begin{table}[htp]
{\small
\begin{center}
\begin{tabular}{c|r r r r r r r r r }
$\bf{(\sigma_2, \mu_2)}$ & \bf{1.0} & \bf{1.5} & \bf{2.0} & \bf{2.5} & \bf{3.0} & \bf{3.5} & \bf{4.0} & \bf{4.5} & \bf{5.0} \\
\hline
{\bf 1.0}   & 2.053  & 2.184  & 2.450  & 2.863  & 3.265  & 3.524  & 3.641  & 3.707  & 3.742  \\
   & {\it(0.477)}  & {\it(0.486)}  & {\it(1.273)}  & {\it(1.085)}  & {\it(0.895)}  & {\it(0.754)}  & {\it(0.706)}  & {\it(0.668)}  & {\it(0.627)}  \\
{\bf 1.5}   & 2.037  & 2.124  & 2.135  & 2.165  & 2.294  & 2.494  & 2.731  & 2.958  & 3.182  \\
   & {\it(0.707)}  & {\it(0.362)}  & {\it(0.404)}  & {\it(0.565)}  & {\it(0.580)}  & {\it(0.668)}  & {\it(0.696)}  & {\it(0.729)}  & {\it(0.706)}  \\
{\bf 2.0}   & 2.026  & 2.067  & 2.121  & 2.059  & 2.083  & 2.132  & 2.259  & 2.546  & 2.807  \\
   & {\it(0.862)}  & {\it(0.337)}  & {\it(0.359)}  & {\it(0.342)}  & {\it(0.350)}  & {\it(0.369)}  & {\it(0.505)}  & {\it(0.648)}  & {\it(0.636)}  \\
{\bf 2.5}   & 1.976  & 2.024  & 2.107  & 2.067  & 2.073  & 2.184  & 2.472  & 2.757  & 2.893  \\
   & {\it(1.027)}  & {\it(0.354)}  & {\it(0.344)}  & {\it(0.302)}  & {\it(0.313)}  & {\it(0.434)}  & {\it(0.570)}  & {\it(0.533)}  & {\it(0.500)}  \\
{\bf 3.0}   & 1.993  & 1.982  & 2.073  & 2.203  & 2.137  & 2.361  & 2.743  & 2.933  & 3.040  \\
   & {\it(1.147)}  & {\it(0.359)}  & {\it(0.329)}  & {\it(0.599)}  & {\it(0.469)}  & {\it(0.580)}  & {\it(0.572)}  & {\it(0.518)}  & {\it(0.486)}  \\
{\bf 3.5}   & 1.979  & 1.964  & 2.076  & 2.186  & 2.367  & 2.662  & 2.973  & 3.181  & 3.354  \\
   & {\it(1.315)}  & {\it(0.432)}  & {\it(0.415)}  & {\it(0.519)}  & {\it(0.938)}  & {\it(0.844)}  & {\it(0.712)}  & {\it(0.678)}  & {\it(0.719)}  \\
{\bf 4.0}   & 1.961  & 1.939  & 2.078  & 2.209  & 2.412  & 2.871  & 3.214  & 3.466  & 3.752  \\
   & {\it(1.469)}  & {\it(0.507)}  & {\it(0.490)}  & {\it(0.500)}  & {\it(0.851)}  & {\it(0.922)}  & {\it(0.807)}  & {\it(0.806)}  & {\it(0.860)}  \\
{\bf 4.5}   & 1.858  & 1.893  & 2.046  & 2.220  & 2.524  & 3.101  & 3.418  & 3.777  & 4.073  \\
   & {\it(1.471)}  & {\it(0.504)}  & {\it(0.445)}  & {\it(0.540)}  & {\it(0.832)}  & {\it(1.034)}  & {\it(0.898)}  & {\it(0.910)}  & {\it(0.901)}  \\
{\bf 5.0}   & 1.698  & 1.886  & 2.063  & 2.273  & 2.678  & 3.187  & 3.673  & 4.046  & 4.295  \\
   & {\it(1.247)}  & {\it(0.605)}  & {\it(0.566)}  & {\it(0.612)}  & {\it(1.128)}  & {\it(1.013)}  & {\it(1.033)}  & {\it(0.970)}  & {\it(0.867)}  \\
\end{tabular}
\end{center}
}
\caption{\label{tab:modfind} 
Average number (and standard deviation) of identified modes on a collection of $1470$ samples.}
\end{table}

The code for both implementing both versions of PMC and for evaluating their mode-finding
abilities was written in {\sf R} \citep{cran} and is available as\\ \verb+http:\\www.ceremade.dauphine.fr\~xian\2RB.R+.
Further documentation is available as\\ \verb+http:\\www.ceremade.dauphine.fr\~xian\2RB.R.pdf+. 

\subsection{Output}\label{sub:compaR}

The experiment was run on $4860$
samples on seven different machines for a total of $238140$ simulations,
using the same machine for both single and double Rao-Blackwellised versions at a given 
value of the parameters in order to keep the CPU comparison sensible. As seen from Table \ref{CPUt}, the CPU time 
required to implement the double Rao-Blackwellised scheme is five to three time higher than the original PMC scheme,
the additional time decreasing with $n$. (This CPU time does not include the mode finding and storing
steps that are required for the comparison. It only corresponds to the execution of a regular PMC run with $10$
iterations.) Note also that the increase
in computing time is certainly less than linear. Both single and double Rao-Blackwellisation PMC samplers
were used on the same samples, with different values of $p$, $\sigma_2$ and $\mu_2$.

\begin{table}[htp]
\begin{center}
\begin{tabular}{r c c}
\bf{$n$\phantom{x}} & \bf{$t_{\rm 1RB}$} & \bf{$t_{\rm 2RB}$} \\
\hline
\hline
 {\bf 20}   &  3.60       &  15.59      \\
      & {\it(0.52)} & {\it(1.12)} \\
 {\bf 30}   &  3.65       &  15.51      \\
      & {\it(0.52)} & {\it(1.13)} \\
 {\bf 40}   &  3.68       &  15.59      \\
      & {\it(0.52)} & {\it(1.13)} \\
 {\bf 50}   &  3.72       &  15.61      \\
      & {\it(0.52)} & {\it(1.13)} \\
 {\bf 100}  &  3.89       &  15.74      \\
      & {\it(0.53)} & {\it(1.16)} \\
 {\bf 500}  &  5.13       &  17.19      \\
      & {\it(0.62)} & {\it(1.26)} \\
 {\bf 1000} &  6.85       &  18.92      \\
           & {\it(0.74)} & {\it(1.42)} \\
\end{tabular}
\end{center}
\caption{\label{CPUt}
Overall mean CPU times and their standard error (in seconds) {\it vs.} sample size. Each 
value is obtained by averaging over $4860$ generated samples, for a range of parameters 
($p=.1,\ldots,.8$, $\mu_2=1,\ldots,5$, $\sigma_2=1,\ldots,5$).}
\end{table}

\begin{figure}
\centerline{\includegraphics[width=10truecm]{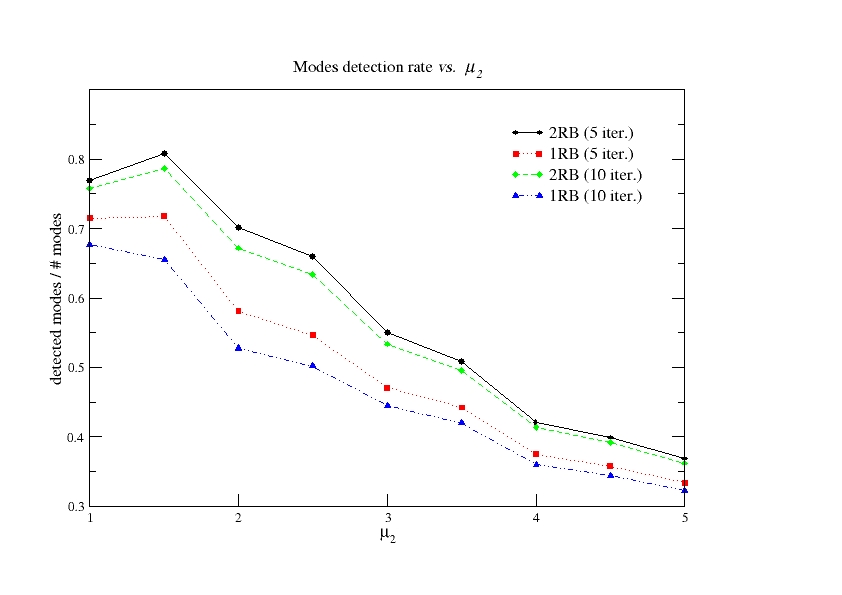}} 
\caption{\label{fig:rate_mu}
Evolution of the capture rate for the single and the double Rao--Blackwellised versions
as a function of $\mu_2$ after $5$ and $10$ iterations, obtained by averaging
Tables \ref{tab_5iter_1RB}--\ref{tab_2RB} over $\sigma_2$.}
\end{figure}

Turning now to the central tables, Tables \ref{tab_5iter_1RB}--\ref{tab_2RB}, we can see that
the influence of $\sigma_2$ on the detection of the modes is relatively limited, in opposition to the
influence of $\mu_2$. As $\mu_2$ increases (recall that $\mu_1=0$), a larger fraction of modes gets 
undetected. Both after $5$ and $10$ iterations of PMC, the performances of the double Rao--Blackwellised scheme
are superior to those of the single Rao--Blackwellisation in terms of mode detection, by a factor of about $5\%$.
Running the PMC algorithm longer clearly has a downward impact on the number of detected modes, even though
this impact is quite limited. It however points out the major issue that importance sampling algorithms like
PMC are mostly unable to recover lost modes. Another interesting feature is that single Rao--Blackwellisation
suffers more from the loss of modes between $5$ and $10$ iterations, compared with double Rao--Blackwellisation.
As expected, the later is more robust because of the average over all past values in the PMC sample.
Figure \ref{fig:rate_mu} presents the evolution of the capture rate for the single and the double
Rao--Blackwellisations  as a function of $\mu_2$ after $5$ and $10$ iterations, obtained by averaging
Tables \ref{tab_5iter_1RB}--\ref{tab_2RB} over $\sigma_2$.

\begin{table}[htp]
{\small
\begin{center}
\begin{tabular}{c r r r r r r r r r }
$\bf{(\sigma_2, \mu_2)}$ & \bf{1.0} & \bf{1.5} & \bf{2.0} & \bf{2.5} & \bf{3.0} & \bf{3.5} & \bf{4.0} & \bf{4.5} & \bf{5.0} \\
\hline
{\bf 1.0}   & 0.743  & 0.714  & 0.631  & 0.536  & 0.452  & 0.409  & 0.399  & 0.386  & 0.387  \\
   & {\it(0.260)}  & {\it(0.256)}  & {\it(0.254)}  & {\it(0.259)}  & {\it(0.232)}  & {\it(0.212)}  & {\it(0.216)}  & {\it(0.210)}  & {\it(0.209)}  \\
{\bf 1.5}   & 0.697  & 0.697  & 0.612  & 0.573  & 0.515  & 0.472  & 0.425  & 0.386  & 0.350  \\
   & {\it(0.272)}  & {\it(0.255)}  & {\it(0.226)}  & {\it(0.208)}  & {\it(0.179)}  & {\it(0.175)}  & {\it(0.157)}  & {\it(0.145)}  & {\it(0.121)}  \\
{\bf 2.0}   & 0.653  & 0.714  & 0.614  & 0.564  & 0.552  & 0.521  & 0.495  & 0.445  & 0.397  \\
   & {\it(0.273)}  & {\it(0.257)}  & {\it(0.226)}  & {\it(0.182)}  & {\it(0.173)}  & {\it(0.151)}  & {\it(0.147)}  & {\it(0.147)}  & {\it(0.131)}  \\
{\bf 2.5}   & 0.666  & 0.718  & 0.602  & 0.563  & 0.540  & 0.509  & 0.452  & 0.394  & 0.378  \\
   & {\it(0.288)}  & {\it(0.256)}  & {\it(0.220)}  & {\it(0.181)}  & {\it(0.155)}  & {\it(0.147)}  & {\it(0.142)}  & {\it(0.114)}  & {\it(0.117)}  \\
{\bf 3.0}   & 0.684  & 0.715  & 0.605  & 0.545  & 0.526  & 0.477  & 0.402  & 0.371  & 0.355  \\
   & {\it(0.302)}  & {\it(0.255)}  & {\it(0.219)}  & {\it(0.193)}  & {\it(0.156)}  & {\it(0.151)}  & {\it(0.119)}  & {\it(0.108)}  & {\it(0.094)}  \\
{\bf 3.5}   & 0.706  & 0.718  & 0.611  & 0.538  & 0.500  & 0.434  & 0.373  & 0.345  & 0.322  \\
   & {\it(0.311)}  & {\it(0.257)}  & {\it(0.225)}  & {\it(0.179)}  & {\it(0.170)}  & {\it(0.152)}  & {\it(0.119)}  & {\it(0.103)}  & {\it(0.083)}  \\
{\bf 4.0}   & 0.725  & 0.719  & 0.610  & 0.535  & 0.484  & 0.409  & 0.347  & 0.318  & 0.291  \\
   & {\it(0.314)}  & {\it(0.258)}  & {\it(0.225)}  & {\it(0.187)}  & {\it(0.155)}  & {\it(0.151)}  & {\it(0.108)}  & {\it(0.094)}  & {\it(0.084)}  \\
{\bf 4.5}   & 0.756  & 0.736  & 0.630  & 0.536  & 0.466  & 0.379  & 0.330  & 0.293  & 0.273  \\
   & {\it(0.307)}  & {\it(0.259)}  & {\it(0.237)}  & {\it(0.192)}  & {\it(0.159)}  & {\it(0.137)}  & {\it(0.111)}  & {\it(0.086)}  & {\it(0.091)}  \\
{\bf 5.0}   & 0.781  & 0.731  & 0.627  & 0.528  & 0.462  & 0.370  & 0.312  & 0.280  & 0.253  \\
   & {\it(0.293)}  & {\it(0.261)}  & {\it(0.238)}  & {\it(0.190)}  & {\it(0.179)}  & {\it(0.136)}  & {\it(0.111)}  & {\it(0.095)}  & {\it(0.079)}  \\
\end{tabular}
\end{center}
}
\caption{\label{tab_5iter_1RB}
Average detection rate after $5$ iterations of the single Rao-Blackwellised PMC
algorithm as a function of $\sigma_2$ and $\mu_2$. 
The number of samples per entry is $1470$, obtained for $7$ values of $n$ and $6$ values of $p$.}
\end{table}

\begin{table}[htp]
{\small
\begin{center}
\begin{tabular}{c r r r r r r r r r }
$\bf{(\sigma_2, \mu_2)}$ & \bf{1.0} & \bf{1.5} & \bf{2.0} & \bf{2.5} & \bf{3.0} & \bf{3.5} & \bf{4.0} & \bf{4.5} & \bf{5.0} \\
\hline
{\bf 1.0}   & 0.794  & 0.801  & 0.725  & 0.618  & 0.528  & 0.474  & 0.463  & 0.438  & 0.440  \\
   & {\it(0.258)}  & {\it(0.247)}  & {\it(0.266)}  & {\it(0.285)}  & {\it(0.265)}  & {\it(0.241)}  & {\it(0.238)}  & {\it(0.232)}  & {\it(0.229)}  \\
{\bf 1.5}   & 0.747  & 0.811  & 0.753  & 0.700  & 0.627  & 0.558  & 0.487  & 0.439  & 0.386  \\
   & {\it(0.275)}  & {\it(0.240)}  & {\it(0.252)}  & {\it(0.257)}  & {\it(0.256)}  & {\it(0.248)}  & {\it(0.223)}  & {\it(0.216)}  & {\it(0.174)}  \\
{\bf 2.0}   & 0.679  & 0.832  & 0.763  & 0.708  & 0.669  & 0.613  & 0.566  & 0.500  & 0.443  \\
   & {\it(0.279)}  & {\it(0.241)}  & {\it(0.255)}  & {\it(0.253)}  & {\it(0.248)}  & {\it(0.232)}  & {\it(0.212)}  & {\it(0.206)}  & {\it(0.191)}  \\
{\bf 2.5}   & 0.679  & 0.832  & 0.756  & 0.697  & 0.648  & 0.586  & 0.514  & 0.447  & 0.416  \\
   & {\it(0.291)}  & {\it(0.240)}  & {\it(0.257)}  & {\it(0.251)}  & {\it(0.238)}  & {\it(0.220)}  & {\it(0.212)}  & {\it(0.185)}  & {\it(0.160)}  \\
{\bf 3.0}   & 0.697  & 0.826  & 0.758  & 0.672  & 0.625  & 0.555  & 0.451  & 0.411  & 0.386  \\
   & {\it(0.304)}  & {\it(0.242)}  & {\it(0.256)}  & {\it(0.263)}  & {\it(0.237)}  & {\it(0.226)}  & {\it(0.179)}  & {\it(0.159)}  & {\it(0.137)}  \\
{\bf 3.5}   & 0.712  & 0.807  & 0.745  & 0.657  & 0.591  & 0.488  & 0.424  & 0.376  & 0.355  \\
   & {\it(0.312)}  & {\it(0.251)}  & {\it(0.260)}  & {\it(0.257)}  & {\it(0.244)}  & {\it(0.212)}  & {\it(0.178)}  & {\it(0.143)}  & {\it(0.131)}  \\
{\bf 4.0}   & 0.728  & 0.788  & 0.740  & 0.646  & 0.571  & 0.457  & 0.386  & 0.350  & 0.317  \\
   & {\it(0.312)}  & {\it(0.258)}  & {\it(0.260)}  & {\it(0.257)}  & {\it(0.238)}  & {\it(0.204)}  & {\it(0.153)}  & {\it(0.138)}  & {\it(0.123)}  \\
{\bf 4.5}   & 0.758  & 0.795  & 0.738  & 0.627  & 0.540  & 0.431  & 0.363  & 0.318  & 0.298  \\
   & {\it(0.307)}  & {\it(0.255)}  & {\it(0.262)}  & {\it(0.249)}  & {\it(0.229)}  & {\it(0.199)}  & {\it(0.146)}  & {\it(0.125)}  & {\it(0.121)}  \\
{\bf 5.0}   & 0.782  & 0.783  & 0.709  & 0.617  & 0.532  & 0.413  & 0.341  & 0.305  & 0.273  \\
   & {\it(0.292)}  & {\it(0.259)}  & {\it(0.264)}  & {\it(0.253)}  & {\it(0.241)}  & {\it(0.182)}  & {\it(0.141)}  & {\it(0.124)}  & {\it(0.105)}  \\
\end{tabular}
\end{center}
}
\caption{\label{tab_5iter_2RB}
Same legend as Table \ref{tab_5iter_1RB} for the double Rao-Blackwellised PMC algorithm.}
\end{table}

\begin{table}[htp]
{\small
\begin{center}
\begin{tabular}{c r r r r r r r r r }
$\bf{(\sigma_2, \mu_2)}$ & \bf{1.0} & \bf{1.5} & \bf{2.0} & \bf{2.5} & \bf{3.0} & \bf{3.5} & \bf{4.0} & \bf{4.5} & \bf{5.0} \\
\hline
{\bf 1.0}   & 0.707  & 0.641  & 0.570  & 0.482  & 0.413  & 0.376  & 0.365  & 0.356  & 0.355  \\
   & {\it(0.259)}  & {\it(0.244)}  & {\it(0.226)}  & {\it(0.220)}  & {\it(0.199)}  & {\it(0.187)}  & {\it(0.187)}  & {\it(0.181)}  & {\it(0.185)}  \\
{\bf 1.5}   & 0.656  & 0.599  & 0.537  & 0.516  & 0.478  & 0.443  & 0.403  & 0.371  & 0.340  \\
   & {\it(0.262)}  & {\it(0.222)}  & {\it(0.170)}  & {\it(0.146)}  & {\it(0.129)}  & {\it(0.129)}  & {\it(0.125)}  & {\it(0.119)}  & {\it(0.102)}  \\
{\bf 2.0}   & 0.638  & 0.622  & 0.533  & 0.518  & 0.512  & 0.491  & 0.474  & 0.427  & 0.384  \\
   & {\it(0.269)}  & {\it(0.231)}  & {\it(0.157)}  & {\it(0.119)}  & {\it(0.115)}  & {\it(0.094)}  & {\it(0.108)}  & {\it(0.115)}  & {\it(0.107)}  \\
{\bf 2.5}   & 0.656  & 0.641  & 0.531  & 0.510  & 0.507  & 0.483  & 0.436  & 0.386  & 0.366  \\
   & {\it(0.285)}  & {\it(0.237)}  & {\it(0.158)}  & {\it(0.111)}  & {\it(0.100)}  & {\it(0.102)}  & {\it(0.117)}  & {\it(0.095)}  & {\it(0.092)}  \\
{\bf 3.0}   & 0.673  & 0.660  & 0.542  & 0.497  & 0.499  & 0.456  & 0.388  & 0.359  & 0.346  \\
   & {\it(0.302)}  & {\it(0.243)}  & {\it(0.168)}  & {\it(0.130)}  & {\it(0.112)}  & {\it(0.114)}  & {\it(0.098)}  & {\it(0.084)}  & {\it(0.080)}  \\
{\bf 3.5}   & 0.700  & 0.660  & 0.554  & 0.499  & 0.472  & 0.416  & 0.362  & 0.335  & 0.315  \\
   & {\it(0.311)}  & {\it(0.245)}  & {\it(0.182)}  & {\it(0.127)}  & {\it(0.122)}  & {\it(0.122)}  & {\it(0.096)}  & {\it(0.085)}  & {\it(0.071)}  \\
{\bf 4.0}   & 0.723  & 0.670  & 0.560  & 0.498  & 0.462  & 0.390  & 0.337  & 0.308  & 0.284  \\
   & {\it(0.312)}  & {\it(0.248)}  & {\it(0.186)}  & {\it(0.139)}  & {\it(0.126)}  & {\it(0.121)}  & {\it(0.095)}  & {\it(0.075)}  & {\it(0.073)}  \\
{\bf 4.5}   & 0.755  & 0.700  & 0.572  & 0.496  & 0.445  & 0.365  & 0.319  & 0.285  & 0.267  \\
   & {\it(0.308)}  & {\it(0.255)}  & {\it(0.201)}  & {\it(0.142)}  & {\it(0.129)}  & {\it(0.126)}  & {\it(0.090)}  & {\it(0.076)}  & {\it(0.079)}  \\
{\bf 5.0}   & 0.780  & 0.700  & 0.578  & 0.501  & 0.438  & 0.357  & 0.303  & 0.270  & 0.249  \\
   & {\it(0.293)}  & {\it(0.258)}  & {\it(0.208)}  & {\it(0.159)}  & {\it(0.149)}  & {\it(0.120)}  & {\it(0.101)}  & {\it(0.078)}  & {\it(0.068)}  \\
\end{tabular}
\end{center}
}
\caption{\label{tab_1RB}
Average detection rate after $10$ iterations of the single Rao-Blackwellised PMC
algorithm as a function of $\sigma_2$ and $\mu_2$.}
\end{table}

\begin{table}[htp]
{\small
\begin{center}
\begin{tabular}{c r r r r r r r r r }
$\bf{(\sigma_2, \mu_2)}$ & \bf{1.0} & \bf{1.5} & \bf{2.0} & \bf{2.5} & \bf{3.0} & \bf{3.5} & \bf{4.0} & \bf{4.5} & \bf{5.0} \\
\hline
{\bf 1.0}   & 0.786  & 0.785  & 0.712  & 0.610  & 0.521  & 0.467  & 0.453  & 0.431  & 0.432  \\
   & {\it(0.259)}  & {\it(0.251)}  & {\it(0.268)}  & {\it(0.284)}  & {\it(0.262)}  & {\it(0.240)}  & {\it(0.236)}  & {\it(0.228)}  & {\it(0.227)}  \\
{\bf 1.5}   & 0.745  & 0.795  & 0.733  & 0.679  & 0.612  & 0.543  & 0.479  & 0.431  & 0.379  \\
   & {\it(0.275)}  & {\it(0.246)}  & {\it(0.253)}  & {\it(0.255)}  & {\it(0.249)}  & {\it(0.240)}  & {\it(0.216)}  & {\it(0.208)}  & {\it(0.167)}  \\
{\bf 2.0}   & 0.677  & 0.813  & 0.736  & 0.684  & 0.647  & 0.600  & 0.555  & 0.494  & 0.437  \\
   & {\it(0.280)}  & {\it(0.246)}  & {\it(0.256)}  & {\it(0.247)}  & {\it(0.242)}  & {\it(0.223)}  & {\it(0.207)}  & {\it(0.200)}  & {\it(0.188)}  \\
{\bf 2.5}   & 0.675  & 0.809  & 0.726  & 0.665  & 0.623  & 0.567  & 0.500  & 0.435  & 0.409  \\
   & {\it(0.291)}  & {\it(0.247)}  & {\it(0.256)}  & {\it(0.244)}  & {\it(0.229)}  & {\it(0.208)}  & {\it(0.199)}  & {\it(0.169)}  & {\it(0.152)}  \\
{\bf 3.0}   & 0.694  & 0.804  & 0.723  & 0.643  & 0.606  & 0.537  & 0.443  & 0.404  & 0.380  \\
   & {\it(0.304)}  & {\it(0.251)}  & {\it(0.257)}  & {\it(0.257)}  & {\it(0.227)}  & {\it(0.211)}  & {\it(0.167)}  & {\it(0.154)}  & {\it(0.134)}  \\
{\bf 3.5}   & 0.711  & 0.783  & 0.706  & 0.624  & 0.568  & 0.477  & 0.413  & 0.372  & 0.350  \\
   & {\it(0.311)}  & {\it(0.257)}  & {\it(0.256)}  & {\it(0.244)}  & {\it(0.233)}  & {\it(0.202)}  & {\it(0.169)}  & {\it(0.135)}  & {\it(0.124)}  \\
{\bf 4.0}   & 0.727  & 0.764  & 0.704  & 0.609  & 0.542  & 0.446  & 0.377  & 0.346  & 0.312  \\
   & {\it(0.313)}  & {\it(0.259)}  & {\it(0.255)}  & {\it(0.241)}  & {\it(0.218)}  & {\it(0.193)}  & {\it(0.145)}  & {\it(0.131)}  & {\it(0.116)}  \\
{\bf 4.5}   & 0.757  & 0.766  & 0.693  & 0.596  & 0.518  & 0.416  & 0.355  & 0.313  & 0.293  \\
   & {\it(0.308)}  & {\it(0.259)}  & {\it(0.258)}  & {\it(0.237)}  & {\it(0.215)}  & {\it(0.185)}  & {\it(0.139)}  & {\it(0.117)}  & {\it(0.116)}  \\
{\bf 5.0}   & 0.782  & 0.764  & 0.665  & 0.589  & 0.512  & 0.404  & 0.334  & 0.300  & 0.268  \\
   & {\it(0.293)}  & {\it(0.260)}  & {\it(0.255)}  & {\it(0.238)}  & {\it(0.228)}  & {\it(0.174)}  & {\it(0.134)}  & {\it(0.120)}  & {\it(0.098)}  \\
\end{tabular}
\end{center}
}
\caption{\label{tab_2RB}
Same legend as Table \ref{tab_1RB} for the double Rao-Blackwellised PMC algorithm.}
\end{table}

\begin{figure}
\centerline{\includegraphics[width=10truecm]{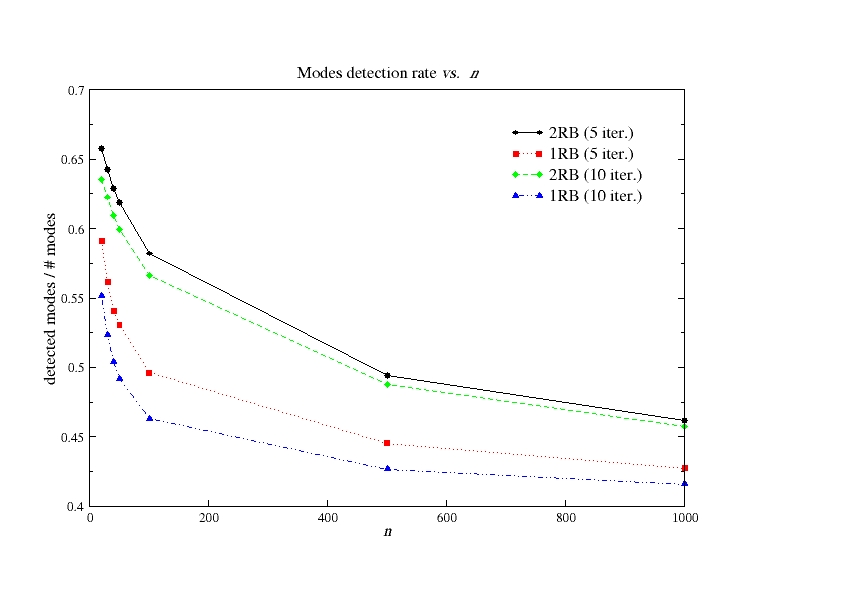}} 
\caption{\label{fig:rate_p}
Evolution of the capture rate for the single and the double Rao--Blackwellisations 
as a function of the sample size $n$ after $5$ and $10$ iterations, obtained by averaging
Tables \ref{tab_5iter_1RB_np}--\ref{tab_2RB_np} over $p$.}
\end{figure}

If we consider instead the output of this simulation experiment decomposed by the values of $n$ and $p$, 
in Tables \ref{tab_5iter_1RB_np}--\ref{tab_2RB_np}, 
a main feature is the quick deterioration in the exploration 
of the modes as $n$ increases. This is obviously
to be expected given that the more observations, the steeper the slopes of the posterior surfaces. Thus,
for small values of $n$, both types of Rao-Blackwellised PMC algorithms achieve high coverage rates, but larger
values of $n$ lead to poorer achievements. Once again, the robustness against the number of PMC iterations is
superior in the case of the double Rao-Blackwellisation. Figure \ref{fig:rate_p} presents the evolution of the capture rate for the single
and the double Rao--Blackwellisations  as a function of the sample size $n$ after $5$ and $10$ iterations, obtained by averaging
Tables \ref{tab_5iter_1RB_np}--\ref{tab_2RB_np} over $p$.

\begin{table}[htp]
{\small
\begin{center}
\begin{tabular}{c r r r r r r r }
$\bf{(p, n)}$ & \bf{20} & \bf{30} & \bf{40} & \bf{50} & \bf{100} & \bf{500} & \bf{1000} \\
\hline
{\bf 0.10}   & 0.602  & 0.565  & 0.542  & 0.519  & 0.478  & 0.423  & 0.411  \\
   & {\it(0.288)}  & {\it(0.276)}  & {\it(0.274)}  & {\it(0.260)}  & {\it(0.243)}  & {\it(0.198)}  & {\it(0.184)}  \\
{\bf 0.20}   & 0.580  & 0.551  & 0.527  & 0.519  & 0.480  & 0.431  & 0.414  \\
   & {\it(0.278)}  & {\it(0.263)}  & {\it(0.249)}  & {\it(0.246)}  & {\it(0.222)}  & {\it(0.193)}  & {\it(0.173)}  \\
{\bf 0.30}   & 0.583  & 0.552  & 0.527  & 0.529  & 0.491  & 0.445  & 0.424  \\
   & {\it(0.276)}  & {\it(0.262)}  & {\it(0.243)}  & {\it(0.246)}  & {\it(0.219)}  & {\it(0.193)}  & {\it(0.167)}  \\
{\bf 0.40}   & 0.582  & 0.558  & 0.538  & 0.530  & 0.496  & 0.455  & 0.432  \\
   & {\it(0.272)}  & {\it(0.259)}  & {\it(0.253)}  & {\it(0.241)}  & {\it(0.221)}  & {\it(0.200)}  & {\it(0.174)}  \\
{\bf 0.50}   & 0.602  & 0.572  & 0.564  & 0.553  & 0.533  & 0.476  & 0.455  \\
   & {\it(0.283)}  & {\it(0.263)}  & {\it(0.256)}  & {\it(0.247)}  & {\it(0.230)}  & {\it(0.192)}  & {\it(0.172)}  \\
{\bf 0.60}   & 0.597  & 0.572  & 0.544  & 0.533  & 0.499  & 0.442  & 0.427  \\
   & {\it(0.276)}  & {\it(0.266)}  & {\it(0.256)}  & {\it(0.242)}  & {\it(0.220)}  & {\it(0.174)}  & {\it(0.153)}  \\
\end{tabular}
\end{center}
}
\caption{\label{tab_5iter_1RB_np}
Average detection rate after $5$ PMC iterations of the single Rao-Blackwellised 
algorithm  as a function of $n$ and $p$. The number 
of samples is $2835$, obtained for $7$ values of 
$\mu_1$ and $5$ values of $\sigma_2$.}
\end{table}

\begin{table}[htp]
{\small
\begin{center}
\begin{tabular}{c r r r r r r r }
$\bf{(p, n)}$ & \bf{20} & \bf{30} & \bf{40} & \bf{50} & \bf{100} & \bf{500} & \bf{1000} \\
\hline
{\bf 0.10}   & 0.652  & 0.619  & 0.601  & 0.577  & 0.542  & 0.463  & 0.445  \\
   & {\it(0.291)}  & {\it(0.286)}  & {\it(0.292)}  & {\it(0.283)}  & {\it(0.283)}  & {\it(0.249)}  & {\it(0.232)}  \\
{\bf 0.20}   & 0.644  & 0.629  & 0.613  & 0.604  & 0.566  & 0.482  & 0.453  \\
   & {\it(0.286)}  & {\it(0.284)}  & {\it(0.283)}  & {\it(0.283)}  & {\it(0.279)}  & {\it(0.250)}  & {\it(0.229)}  \\
{\bf 0.30}   & 0.653  & 0.645  & 0.630  & 0.628  & 0.588  & 0.500  & 0.467  \\
   & {\it(0.288)}  & {\it(0.288)}  & {\it(0.285)}  & {\it(0.284)}  & {\it(0.280)}  & {\it(0.256)}  & {\it(0.228)}  \\
{\bf 0.40}   & 0.664  & 0.662  & 0.654  & 0.643  & 0.598  & 0.513  & 0.472  \\
   & {\it(0.291)}  & {\it(0.289)}  & {\it(0.290)}  & {\it(0.281)}  & {\it(0.283)}  & {\it(0.260)}  & {\it(0.227)}  \\
{\bf 0.50}   & 0.668  & 0.651  & 0.648  & 0.639  & 0.614  & 0.524  & 0.482  \\
   & {\it(0.298)}  & {\it(0.290)}  & {\it(0.290)}  & {\it(0.284)}  & {\it(0.275)}  & {\it(0.241)}  & {\it(0.211)}  \\
{\bf 0.60}   & 0.665  & 0.651  & 0.627  & 0.623  & 0.582  & 0.482  & 0.450  \\
   & {\it(0.291)}  & {\it(0.288)}  & {\it(0.285)}  & {\it(0.276)}  & {\it(0.272)}  & {\it(0.222)}  & {\it(0.189)}  \\
\end{tabular}
\end{center}
}
\caption{\label{tab_5iter_2RB_np}
Same legend as Table \ref{tab_5iter_1RB_np} in the double Rao-Blackwellised case.}
\end{table}

\begin{table}[htp]
{\small
\begin{center}
\begin{tabular}{c r r r r r r r }
$\bf{(p, n)}$ & \bf{20} & \bf{30} & \bf{40} & \bf{50} & \bf{100} & \bf{500} & \bf{1000} \\
\hline
{\bf 0.10}   & 0.577  & 0.541  & 0.519  & 0.495  & 0.457  & 0.408  & 0.398  \\
   & {\it(0.281)}  & {\it(0.266)}  & {\it(0.259)}  & {\it(0.243)}  & {\it(0.223)}  & {\it(0.175)}  & {\it(0.161)}  \\
{\bf 0.20}   & 0.545  & 0.514  & 0.492  & 0.485  & 0.447  & 0.413  & 0.402  \\
   & {\it(0.265)}  & {\it(0.242)}  & {\it(0.222)}  & {\it(0.218)}  & {\it(0.183)}  & {\it(0.164)}  & {\it(0.152)}  \\
{\bf 0.30}   & 0.541  & 0.510  & 0.489  & 0.480  & 0.456  & 0.424  & 0.414  \\
   & {\it(0.257)}  & {\it(0.233)}  & {\it(0.214)}  & {\it(0.201)}  & {\it(0.180)}  & {\it(0.159)}  & {\it(0.146)}  \\
{\bf 0.40}   & 0.536  & 0.516  & 0.496  & 0.487  & 0.458  & 0.431  & 0.418  \\
   & {\it(0.252)}  & {\it(0.234)}  & {\it(0.219)}  & {\it(0.207)}  & {\it(0.178)}  & {\it(0.162)}  & {\it(0.143)}  \\
{\bf 0.50}   & 0.564  & 0.542  & 0.528  & 0.517  & 0.497  & 0.459  & 0.443  \\
   & {\it(0.269)}  & {\it(0.246)}  & {\it(0.232)}  & {\it(0.219)}  & {\it(0.197)}  & {\it(0.166)}  & {\it(0.156)}  \\
{\bf 0.60}   & 0.544  & 0.519  & 0.499  & 0.487  & 0.463  & 0.426  & 0.420  \\
   & {\it(0.258)}  & {\it(0.238)}  & {\it(0.224)}  & {\it(0.206)}  & {\it(0.180)}  & {\it(0.146)}  & {\it(0.134)}  \\
\end{tabular}
\end{center}
}
\caption{\label{tab_1RB_np}
Average detection rate after $10$ PMC iterations for the single
Rao--Blackwellised algorithm as a function of $n$ and $p$.}
\end{table}

\begin{table}[htp]
{\small
\begin{center}
\begin{tabular}{c r r r r r r r }
$\bf{(p, n)}$ & \bf{20} & \bf{30} & \bf{40} & \bf{50} & \bf{100} & \bf{500} & \bf{1000} \\
\hline
{\bf 0.10}   & 0.639  & 0.608  & 0.590  & 0.564  & 0.529  & 0.458  & 0.441  \\
   & {\it(0.291)}  & {\it(0.285)}  & {\it(0.288)}  & {\it(0.277)}  & {\it(0.275)}  & {\it(0.243)}  & {\it(0.227)}  \\
{\bf 0.20}   & 0.622  & 0.612  & 0.594  & 0.585  & 0.549  & 0.475  & 0.448  \\
   & {\it(0.284)}  & {\it(0.281)}  & {\it(0.278)}  & {\it(0.276)}  & {\it(0.267)}  & {\it(0.246)}  & {\it(0.220)}  \\
{\bf 0.30}   & 0.629  & 0.623  & 0.608  & 0.604  & 0.571  & 0.493  & 0.461  \\
   & {\it(0.284)}  & {\it(0.283)}  & {\it(0.279)}  & {\it(0.277)}  & {\it(0.274)}  & {\it(0.250)}  & {\it(0.223)}  \\
{\bf 0.40}   & 0.641  & 0.636  & 0.628  & 0.620  & 0.582  & 0.506  & 0.468  \\
   & {\it(0.287)}  & {\it(0.285)}  & {\it(0.286)}  & {\it(0.276)}  & {\it(0.274)}  & {\it(0.254)}  & {\it(0.224)}  \\
{\bf 0.50}   & 0.642  & 0.630  & 0.627  & 0.619  & 0.600  & 0.518  & 0.480  \\
   & {\it(0.293)}  & {\it(0.284)}  & {\it(0.283)}  & {\it(0.277)}  & {\it(0.269)}  & {\it(0.237)}  & {\it(0.208)}  \\
{\bf 0.60}   & 0.641  & 0.628  & 0.608  & 0.604  & 0.569  & 0.475  & 0.447  \\
   & {\it(0.287)}  & {\it(0.281)}  & {\it(0.278)}  & {\it(0.271)}  & {\it(0.264)}  & {\it(0.215)}  & {\it(0.184)}  \\
\end{tabular}
\end{center}
}
\caption{\label{tab_2RB_np}
Same legend as Table \ref{tab_1RB_np} for the double Rao--Blackwellised PMC algorithm.}
\end{table}

\begin{figure}
\begin{center}
\includegraphics[width=6truecm]{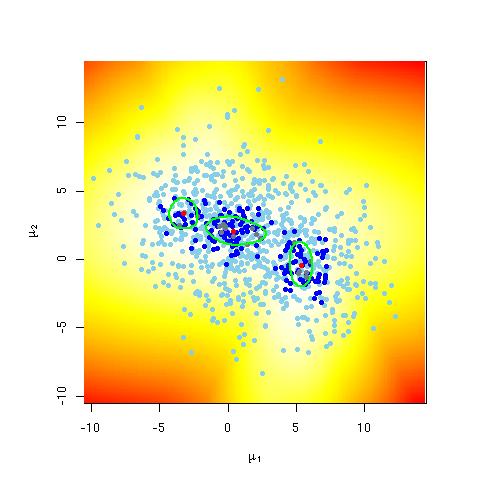} 
\includegraphics[width=6truecm]{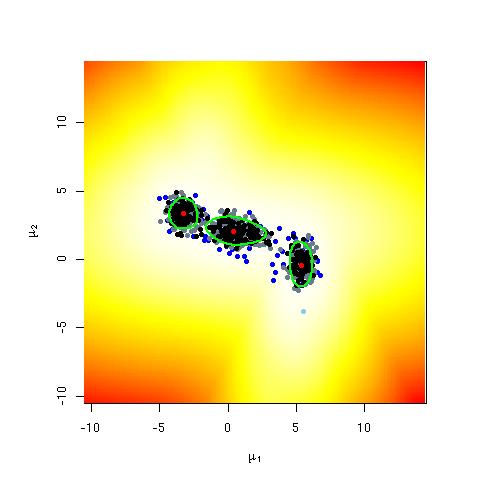} 

\includegraphics[width=6truecm]{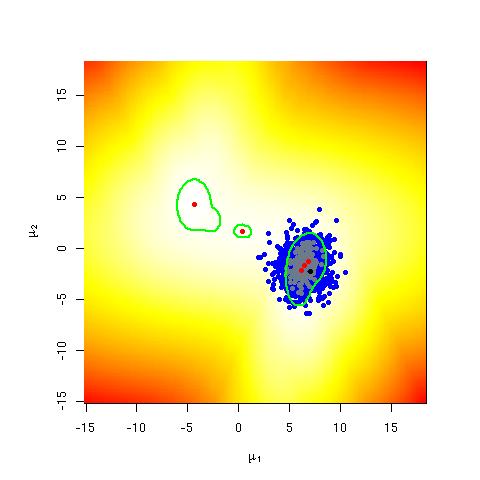} 
\includegraphics[width=6truecm]{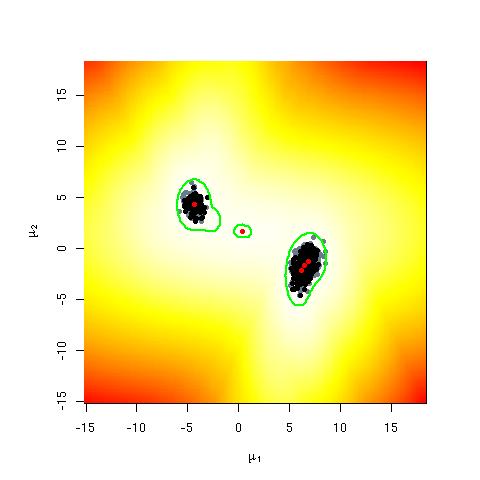} 

\includegraphics[width=6truecm]{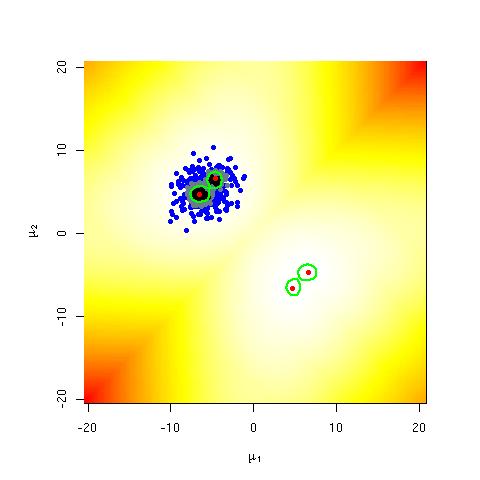} 
\includegraphics[width=6truecm]{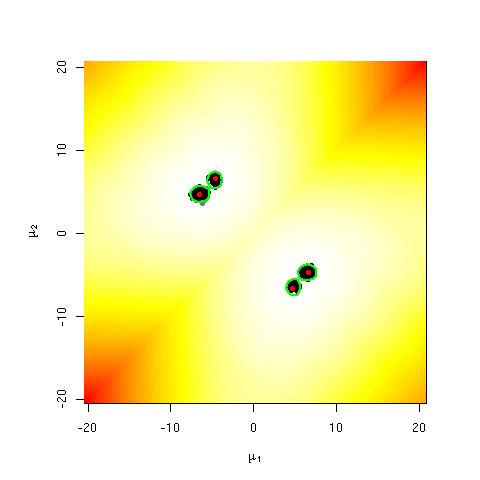} 

\caption{\label{fig:comparison}
Comparison between PMC samples using a single Rao-Blackwellisation (left) and a double Rao-Blackwellisation (right).}

\end{center}
\end{figure}

Figure \ref{fig:comparison} illustrates the superior performances of
the double Rao-Blackwellisation strategy, compared to the single one in terms of modes detection.
Those three graphs plot side by side for a given sample the repartition of the PMC samples in the modal
domains. The colour codes are associated with the five possible variances used in the $5$ kernel
proposal. In those three experiments, the persistence of the samples in all modes, as well as the concentration
in the regions of importance, is noticeable. Both rows of Figure \ref{fig:comparison} interestingly 
contain a ridge structure in one of the modal basins.

\section{Conclusions}\label{sec:concq}
The double Rao-Blackwellised algorithm provides a more robust framework for adapting
general importance sampling densities represented as mixtures in the sense that the
influence of the starting points diminishes when compared with the original algorithm. 
It is thus unnecessary to rely on a large value of the number $T$
of iterations of the PMC algorithm: with large enough sample sizes $N$
at each iteration---especially on the initial iteration that requires many
points to counter-weight a potentially poor initial proposal---, it is quite
uncommon to fail to spot a stabilisation of both the estimates and of the
entropy criterion within a few iterations.

\end{document}